\documentclass[twocolumn:]{pasj02}
\usepackage[switch,mathlines]{lineno} 

\Received{}
\Accepted{}
 
\usepackage{amsmath}

\usepackage[whole]{bxcjkjatype}
\usepackage{url}
\usepackage[T1]{fontenc}
\usepackage{appendix}
\usepackage{textcomp}  
\usepackage{comment}
\usepackage{ulem}
\usepackage{subcaption}

\begin{document}

\title{Exploring the dynamics of the Coma galaxy cluster by mapping its X-ray emission line profiles with XRISM}

\author{
Kosei \textsc{Sakai}\altaffilmark{1},
Kazuhiro \textsc{Nakazawa}*,\altaffilmark{2,1} \email{nakazawa@u.phys.nagoya-u.ac.jp}
\orcid{0000-0003-2930-350X}
Maxim \textsc{Markevitch},\altaffilmark{3} \orcid{0000-0003-0144-4052}
Dominique \textsc{Eckert},\altaffilmark{4} \orcid{0000-0001-7917-3892}
Yuichiro \textsc{Ezoe},\altaffilmark{5} 
Yuto \textsc{Ichinohe},\altaffilmark{6} \orcid{0000-0002-6102-1441}
Richard \textsc{Kelley},\altaffilmark{3} \orcid{0009-0007-2283-3336}
Richard \textsc{Mushotzky},\altaffilmark{3} \orcid{0000-0002-7962-5446}
Nobuhiro \textsc{Okabe},\altaffilmark{7} \orcid{0000-0003-2898-0728}
Yuki \textsc{Omiya},\altaffilmark{8} \orcid{0009-0009-9196-4174}
Naomi \textsc{Ota},\altaffilmark{9} \orcid{0000-0002-2784-3652}
Cicely \textsc{Potter},\altaffilmark{10} \orcid{0009-0003-7738-9173}
Andrew \textsc{Szymkowiak},\altaffilmark{11} 
Shunsuke \textsc{Torii},\altaffilmark{1} 
Nhut \textsc{Truong} ,\altaffilmark{12,3, 13} \orcid{0000-0003-4983-0462}
Ay\c{s}eg\"{u}l T\"{u}mer, \altaffilmark{12,3,13} \orcid{0000-0002-3132-8776} 
Yuusuke \textsc{Uchida},\altaffilmark{14} \orcid{0000-0002-7962-4136}
Dan \textsc{Wik},\altaffilmark{\color{black} 10} \orcid{0000-0001-9110-2245}
Irina \textsc{Zhuravleva},\altaffilmark{15} \orcid{0000-0001-7630-8085} 
and
John \textsc{ZuHone}\altaffilmark{16} \orcid{0000-0003-3175-2347}
}

\altaffiltext{1}{Graduate School of Science, Nagoya University, Nagoya, Aichi 464-8602, Japan}
\altaffiltext{2}{Kobayashi-Maskawa Institute for the Origin of Particles and the Universe, Nagoya University, Nagoya, Aichi 464-8601, Japan}
\altaffiltext{3}{NASA / Goddard Space Flight Center, Greenbelt, MD 20771, USA}
\altaffiltext{4}{Department of Astronomy, University of Geneva, Versoix CH-1290, Switzerland}
\altaffiltext{5}{Department of Physics, Tokyo Metropolitan University, Tokyo 192-0397, Japan}
\altaffiltext{6}{RIKEN Nishina Center, Saitama 351-0198, Japan}
\altaffiltext{7}{Graduate School of Advanced Science and Engineering, Hiroshima University, Higashi-Hiroshima, Hiroshima 739-8526, Japan}
\altaffiltext{8}{Institute of Space and Astronautical Science (ISAS), Japan Aerospace Exploration Agency (JAXA), Kanagawa 252-5210, Japan}
\altaffiltext{9}{Department of Physics, Nara Women's University, Nara 630-8506, Japan}
\altaffiltext{10}{The University of Utah, Salt Lake City, UT 84112, USA}
\altaffiltext{11}{Yale Center for Astronomy and Astrophysics, Yale University, CT 06520-8121, USA}
\altaffiltext{12}{Center for Space Sciences and Technology, University of Maryland, Baltimore County (UMBC), Baltimore, MD, 21250 USA}
\altaffiltext{13}{Center for Research and Exploration in Space Science and Technology, NASA / GSFC (CRESST II), Greenbelt, MD 20771, USA}
\altaffiltext{14}{\color{black} The University of Tokyo, Bunkyo-ku, Tokyo 113-0033, Japan \color{black}}

\altaffiltext{15}{Department of Astronomy and Astrophysics, University of Chicago, Chicago, IL 60637, USA}
\altaffiltext{16}{Center for Astrophysics, Harvard \& Smithsonian, Cambridge, MA 02138, USA}


\maketitle

\begin{abstract}

The intracluster medium (ICM) in merging galaxy clusters exhibits turbulence and bulk flows. Unraveling these components is crucial not only for elucidating the geometry of the cluster mergers, but also for understanding the physics of magnetic field amplification and relativistic particle acceleration. 
XRISM/Resolve data for two $3'\times3'$ fields in the core of the Coma cluster reveales that the ICM in the central field moves with $\Delta cz = -430$~km~s$^{-1}$ relative to the cluster galaxy average, while 
that in the southern field moves with $\Delta cz = -730$~km~s$^{-1}$ (see \cite{2025ApJ...985L..20X}, hereinafter ``Paper I''). 
In this paper, we perform a more detailed analysis of these data sets to search for non-Gaussian features in the \color{black} Fe-K line complex \color{black}
profiles. In the spectra from the northwest (NW) quadrant of the central field, in addition to the 
main and
redshifted ICM components 
\color{black} ($\Delta cz = -6$ km s$^{-1}$) \color{black}
reported in Paper I, we find evidence of 
another, blueshifted component, moving with 
\color{black}
$\Delta cz = -1274$ km s$^{-1}$.
\color{black}
For a systematic search for other significant velocity components, we perform a bias-free 3 eV step multi-component fit to the Resolve full-array spectra from the central and southern fields. This search uncovers another redshifted component in the southern field, moving with 
\color{black}
$\Delta cz \sim +1250$ km s$^{-1}$. 
\color{black}
We estimate the energy densities of the ICM turbulence and bulk motion to be similar to each other and several times greater than the energy density of the cluster's $B\sim 5~\mu$G magnetic field.
\end{abstract}

\section{Introduction}

Mergers of clusters of galaxies drive large-scale motions in the intracluster medium (ICM) on Mpc scales. Such events release up to $10^{64}$~erg of gravitational energy, a significant fraction of which is transferred to the ICM. X-ray observations reveal distorted ICM morphologies, including shock fronts, and hot gas produced by the conversion of bulk kinetic energy into thermal energy (\cite{1999ApJ...521..526M}). In merging clusters, synchrotron radio halos extending across the cluster and radio relics at the periphery are commonly observed in the 100 MHz\UTF{2013}GHz range. These features are considered evidence of GeV electron acceleration in magnetic fields of a few $\mu$G (\cite{2014IJMPD..2330007B,2019SSRv..215...16V}).
Understanding the energy release associated with cluster mergers requires knowledge of the merger geometry and the overall ICM velocity field, including line-of-sight velocity dispersions - measurements that were difficult to obtain with X-ray CCD instruments.


The Coma cluster (Abell 1656) is a well-known nearby merging system, hosting $\sim 1000$ member galaxies and an average redshift of $z = 0.02333$ ($cz = 6995 \pm 37$~km~s$^{-1}$; \cite{10.1093/mnras/sty2379}). 
The center hosts two giant elliptical galaxies : NGC 4889 to the east ($cz = 6446 \pm 3$~km~s$^{-1}$) and NGC 4874 to the west ($cz = 7167 \pm 2$~km~s$^{-1}$). Relative to the cluster mean, NGC 4889 is blueshifted by $\Delta cz = -549$~km~s$^{-1}$, while NGC 4874 is redshifted by 
$\Delta cz = 172$~km~s$^{-1}$. 
The ICM is hot, reaching $\sim 11$~keV in the northwest and $\sim 6$~keV in the southeast (\cite{2020AandA...633A..42S}). Using gain recalibration based on the Cu K-line in XMM-Newton/PN data, \citet{2020AandA...633A..42S} reported large-scale bulk velocity patterns: redshifted by $\Delta cz = +200$--$300$~km~s$^{-1}$ in the north and blueshifted by $\Delta cz = -200$~km~s$^{-1}$ in the south,
\color{black} while the systematic error for the analyses was estimated to be $\pm 150$~km~s$^{-1}$. \color{black}

The X-ray morphology of the ICM is broadly 
spherically symmetric but exhibits local structures, including an eastern excess \color{black} located $\sim 750$~kpc from NGC~4874 \color{black} (\cite{1997ApJ...474L...7V,Neumann03}), the infalling NGC 4839 group located \color{black} 1.6~Mpc \color{black} to the south west with $\Delta cz = 340$~km~s$^{-1}$ (\cite{1994ApJ...427L..87B, 2001AandA...365L..74N,2019MNRAS.485.2922L, 2020AandA...634A..30M,2020MNRAS.497.3204M}), and a shock front concentric with the cluster in the north and west outskirts \color{black} at a radius of $\sim 2.5$~Mpc \color{black} (\cite{Churazov21}). The cluster also hosts one of the brightest radio halos ($\sim 0.5$~Jy at 1.4 GHz; \cite{1997AandA...321...55D}), covering the central $\sim 500$~kpc and roughly tracing the ICM distribution. Recent LOFAR 144~MHz mapping reveales a non-circular halo with a bridge connecting the NGC~4839 group, the radio relic, and the radio galaxy NGC~4789; its northern and western edges 
\color{black}
of the halo
\color{black}
coincide with an ICM shock (\cite{Bonafede_2022}). Rotation measure studies estimate a magnetic field of $\sim 4.7~\mu$G within the core radius (\cite{2010AandA...513A..30B}),
consistent with the lack of strong inverse Compton hard X-ray emission (e.g. Suzaku/HXD; \cite{2009ApJ...696.1700W}, and Swift/BAT; \cite{2011ApJ...727..119W}).

These observational results indicate that the Coma cluster is actively merging. However, the overall optical redshift distribution remains flat as a function of radius (e.g. see Figure~9 of \cite{2021AandA...650A..76H}), and the two-dimensional redshift map shows no strong global bias aside from the NGC 4839 group (Figure~2 of \cite{2021AandA...650A..76H}). In contrast, X-ray temperature and velocity maps reveal north\UTF{2013}south deviations (\cite{2020AandA...633A..42S}), while the two brightest cluster galaxies are aligned east\UTF{2013}west. Consequently, the merger axis, phase, and geometry remain uncertain.

\begin{figure}
\centering
\includegraphics[width=8cm]{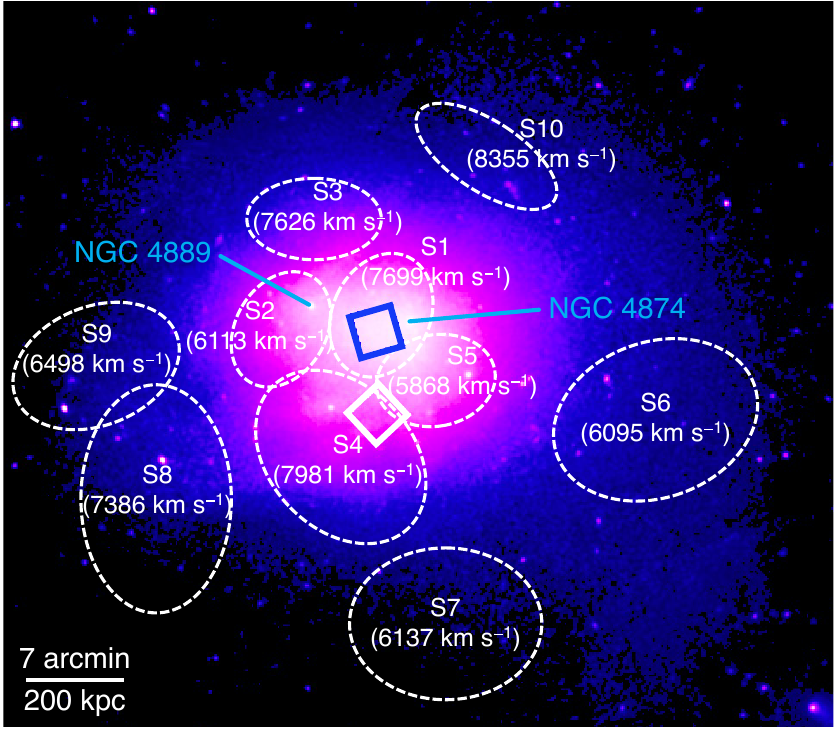}
\vspace*{0.1cm}
\caption{Pointing location of the two XRISM PV Coma cluster observations, overlaid on an XMM-Newton/PN X-ray image in the 0.7--12~keV band. The blue square denotes the Center observation and the white square denotes the South observation. The two BCGs are marked with blue \color{black} labels. \color{black} 
Galaxy groups identified from optical redshifts (\cite{2021AandA...650A..76H}) lie within the white dashed regions, and the accompanying numbers indicate their bulk velocities. {Alt text: Image showing the X-ray surface brightness in the 0.7 to 12 kilo electron volt band.}}
\label{fig:image}
\end{figure}

XRISM (\cite{2025PASJ...77S...1T}) observed the Coma cluster during its performance verification (PV) phase with two pointings near the center as shown in Figure~\ref{fig:image} (see also Section~\ref{sec:obs}). The PV observations with the Resolve instrument 
\color{black}
(\cite{2025JATIS..11d2023I,2025JATIS..11d2026K}) 
\color{black}
reveales relatively low velocity dispersions $\sigma_{cz} = 200$--$230$~km~s$^{-1}$ ($1 \sigma$) and a large bulk velocity difference between the two pointings (\cite{2025ApJ...985L..20X}, hereafter Paper I). The ICM around the X-ray peak (hereafter \texttt{Center}) is blueshifted by $\Delta cz = -430$~km~s$^{-1}$, and 
the region located $\sim 6'$ south of \texttt{Center} (hereafter \texttt{South}) is blueshifted by $-730$~km~s$^{-1}$ relative to the optical mean. The third observation during the first General Observer program located $\sim6'$ north of \texttt{Center} is not discussed here, but shows similar trend, with $\Delta cz = -199 \pm26$~km~s$^{-1}$ and $\sigma_{cz} = 167 \pm 39$~km~s$^{-1}$ (\cite{2026AandA...706L..21G}). These results indicate that bulk motions dominate over turbulence.

The 
\color{black} Fe He-$\alpha$ \color{black} line profile in the \texttt{Center} spectra is not well 
modeled by a single Gaussian; instead, it exhibits a redshifted shoulder, particularly in the northwestern (NW) quadrant of the Resolve field of view (FoV; see Figure 4 of Paper I). Theoretical studies predict that ICM emission lines may exhibit multi-component profiles, i.e., showing a non-Gaussian profile shape (\cite{2003AstL...29..791I}). They also suggest that the most favorable energy range for detecting hydrodynamic broadening effects lies toward the blueshifted tail of the \color{black} Fe He-$\alpha$ \color{black} complex.

Motivated by these findings, we adopt two approaches to search for non-Gaussian 
profiles \color{black} in Fe-K line complex. \color{black}
Firstly, we apply a multiple Gaussian velocity dispersion model fit to the NW-quadrant spectra of \texttt{Center}, as well as the full-array, to 
assess the significance of the second Gaussian component,
and also search for a third component.
Secondly, we apply a multi-component fitting method using fixed redshift steps of 3 eV -- slightly smaller than the Resolve energy resolution (
4.5 eV full width at half maximum (FWHM) at 6 keV, \cite{2025PASJ...77S...1T}
) -- to identify red/blueshifted components 
assuming solely thermal broadening.
Throughout this paper, we assume a flat cosmology with $H_0=70$ km~s$^{-1}$~Mpc$^{-1}$ and $\Omega_m=0.3$, corresponding to $1'=28.2$~kpc at the cluster redshift. All errors are quoted at the 1$\sigma$ level unless otherwise noted, and all velocities and redshifts are given in the \color{black} barycentric \color{black} frame.

\section{XRISM Observation and Data analysis}
\label{sec:obs}

The 
\texttt{Center} observation (ObsID: 300073010) targeted ($\alpha = 194.944 \degree$,
$\delta = 27.947\degree$), near the cluster's X-ray surface brightness peak. 
Two additional observations (ObsIDs: 300074010 and 300074020) were performed at ($\alpha = 194.941 \degree$, $\delta = 27.847\degree$),  $6'$ south of \texttt{Center}. Combined, they are referred to as \texttt{South}.
\color{black}
Barycentric velocity corrections are applied, estimated using \texttt{barycen}, which are $-23.5$~km~s$^{-1}$ (\texttt{Center}) and $-21.6$~km~s$^{-1}$ (\texttt{South}). For simplicity, we fitted all spectra without barycentric correction, and simply modified the fitted value with these offset velocities.
\color{black}

Details of the data screening are described in paper I, and here we provide a brief summary.
The Resolve data were reprocessed with the XRISM pipeline (Build 8) with CalDB version 8.
The effective exposure times after screening are 398~ks for \texttt{Center}, and a total of 158~ks for \texttt{South} (85~ks for 300074010 and 73~ks for 300074020). 
In all analyses, we excluded Resolve pixel 27, which exhibited unstable gain behavior, and pixel 12, designated for calibration purposes. Only high-resolution primary events (``Hp'' or \verb|ITYPE = 0|) are used, 
\color{black}
and therefore the CalDB version (8) we used in this study (as well as Paper I) shows little difference to the newest one (13) within 0.01\% in Resolve gain.
\color{black}
Instrument responses were generated with Build 8 software. The redistribution matrix file (RMF) was configured to use the ``L'' size\footnote{\url{heasarc.gsfc.nasa.gov/docs/xrism/analysis/abc_guide/xrism_abc.html}}, and the ancillary response file (ARF) was produced with \verb|xaarfgen|, assuming a point source located at the center of each region.
The non-X-ray background (NXB) spectrum was derived from Resolve night-Earth data using \verb|rslnxbgen|, with weights applied based on the geomagnetic cutoff rigidity distribution observed during each observation.
The NXB spectra were modeled with a power-law and Gaussian lines, and the best-fit model was incorporated as the NXB component in all spectral fitting \footnote{\url{https://heasarc.gsfc.nasa.gov/docs/xrism/analysis/nxb/index.html}}.
The cosmic X-ray background (CXB) contributes less than $<1 \%$ of the cluster signal in this band and was therefore ignored. 

\section{Spectral modeling with multiple Gaussian velocity dispersions}
\label{S_2kt}

\subsection{One component fit}

\begin{figure}[h]
\centering
\includegraphics[width=8cm]{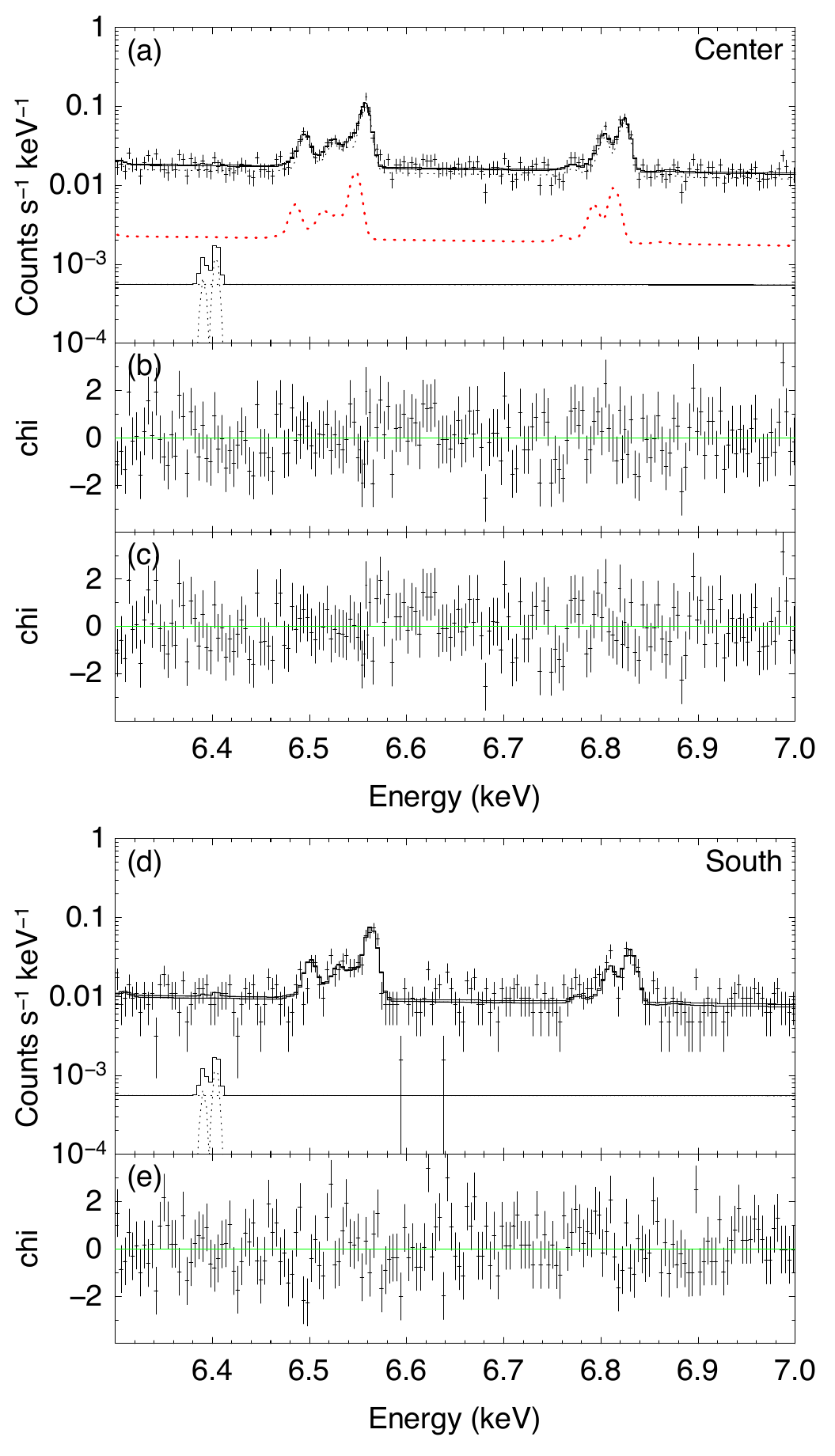}
\caption{
(a) Close-up around Fe-K line \color{black} complex \color{black} regions of the center pointing, fitted with the \color{black} two components  \color{black}  model. (b) Residual plot for the \color{black} one component model, and (c) that of the two components  \color{black}  model. (d) Spectra of the south pointing, and (e) residual plot for the \color{black} one component  \color{black}  model. 
{Alt text: 
\color{black} Five 
\color{black}
line graphs. In all panels (a)--(e), the x axis shows the energy from 6.3 to 7 kilo electron volt. In panels (a) and (d), the y axis shows the count from \color{black} 0.0001 to 1 \color{black} counts per second and per kilo electron volt. In panels (b), (c), and (e), the residual from minus 4 to 4.}}
\label{fig:spec1}
\end{figure}

To investigate 
ICM motions, we first fit the XRISM/Resolve spectra with a thermal model incorporating Gaussian velocity dispersion, using the \texttt{bapec} model in XSPEC. 
In addition to temperature, metal abundance, and normalization, the mean redshift and Gaussian-modeled velocity dispersion were treated as free parameters 
\color{black}
(hereafter, one component (or 1 comp) model).
\color{black}
The absorption column density was fixed at $N_{\rm{H}} =9.2\times 10^{19}~\rm{cm}^{-2}$, following Paper I. 
We applied the 
\color{black}
one component 
\color{black}
model to the wide-band 2--10~keV spectrum of the \texttt{Center} observation integrated over the full array. A close-up of the Fe-K band 
\color{black}
is shown in Figure~\ref{fig:spec1} (a),
and the residuals from  the one component fit is shown in Figure~\ref{fig:spec1} (b). \color{black}
The same procedure was applied to the \texttt{South} spectrum, as shown in Figure~\ref{fig:spec1} (d) and (e).

The results are summarized in 
\color{black}
the top third of 
\color{black}
Table~\ref{table:1ktfit} and are consistent with those reported in Paper I, which uses the 2--9~keV band. 
The ICM in the \texttt{Center} and \texttt{South} regions are blueshifted by about $\Delta cz = -450$~km~s$^{-1}$ and $-750$~km~s$^{-1}$,
respectively, relative to the cluster optical mean, and the velocity dispersions are both $\sigma_{cz} = 217$~km~s$^{-1}$.

\begin{table}[htbp]
\tbl{Fit parameters of the \color{black} one, two and three components \color{black} model 
}{%
\begin{tabular}{llll}  
\hline\hline\noalign{\vskip1pt} 
\multicolumn{4}{c}{One component model fit to the two full-array spectra}\\
\hline
Region & Center & \multicolumn{2}{l}{~~~South}  \\
\hline
$kT$ [keV] & $8.36^{+0.14}_{-0.14}$ & \multicolumn{2}{l}{~~~$7.44^{+0.24}_{-0.24}$} \\
$Z$ [solar]$^1$ & $0.32^{+0.01}_{-0.01}$ & \multicolumn{2}{l}{~~~$0.36^{+0.03}_{-0.02}$} \\
$\Delta cz$ [km s$^{-1}$] $^2$  & $-450^{+12}_{-13}$ & \multicolumn{2}{l}{~~~$-744^{+22}_{-23}$} \\
$\sigma_{cz}$ [km s$^{-1}$] $^3$  & $217^{+13}_{-13}$ & \multicolumn{2}{l}{~~~$217^{+24}_{-22}$} \\
{\it Norm} [$\times 10^{-3}$ cm$^{-5}$] $^4$  & $8.93^{+0.08}_{-0.08}$ & \multicolumn{2}{l}{~~~$5.13^{+0.10}_{-0.10}$} \\ 
{C-stat}/{d.o.f} & $17345/15994$ & \multicolumn{2}{l}{~~~$16302/15994$}  \\ 
\hline\hline
\noalign{\vskip12pt} 
%
\hline\hline
\noalign{\vskip1pt} 
\multicolumn{4}{c}{Two components model fit to Center NW quadrant}\\
\hline
component & 1st  & \multicolumn{2}{l}{~~~2nd} \\
\hline
$kT$ [keV] & $8.48^{+0.28}_{-0.27}$ & \multicolumn{2}{l}{~~~--} \\
$Z$ [solar] & $0.32^{+0.02}_{-0.02}$ & \multicolumn{2}{l}{~~~--} \\
$\Delta cz$ [km s$^{-1}$]  & $-498^{+31}_{-33}$  & \multicolumn{2}{l}{~~~$5^{+101}_{-94}$} \\
$\sigma_{cz}$ [km s$^{-1}$]  & $124^{+44}_{-53}$ & \multicolumn{2}{l}{~~~--} \\
{\it Norm} [$\times 10^{-3}$ cm$^{-5}$] & $1.63^{+0.15}_{-0.15}$ & \multicolumn{2}{l}{~~~$0.57^{+0.15}_{-0.15}$} \\ 
{C-stat}/{d.o.f} & \multicolumn{2}{c}{~~~$16755/15992$}  & \\
\hline\noalign{\vskip3pt} 
\multicolumn{4}{c}{Two components model fit to Center full-array}\\
\hline
component & 1st  & \multicolumn{2}{l}{~~~2nd} \\
\hline
$kT$ [keV] & $8.36^{+0.14}_{-0.14}$ & \multicolumn{2}{l}{~~~--} \\
$Z$ [solar] & $0.33^{+0.01}_{-0.01}$ & \multicolumn{2}{l}{~~~--} \\
$\Delta cz$ [km s$^{-1}$]  & $-481^{+25}_{-38}$ & \multicolumn{2}{l}{~~~$-6^{+229}_{-151}$}  \\
$\sigma_{cz}$ [km s$^{-1}$]  & $177^{+27}_{-32}$ & \multicolumn{2}{l}{~~~--}\\
{\it Norm} [$\times 10^{-3}$ cm$^{-5}$] & $7.80^{+0.60}_{-1.01}$ & \multicolumn{2}{l}{~~~$1.12^{+1.01}_{-0.59}$} \\ 
{C-stat}/{d.o.f} & \multicolumn{2}{c}{$17338/15992$}  & \\
\hline\hline\noalign{\vskip12pt} 
\hline\hline\noalign{\vskip3pt} 
\multicolumn{4}{c}{Three components model fit to Center NW quadrant}\\
\hline
component & 1st  & 2nd  & 3rd \\
\hline
$kT$ [keV] & $8.38^{+0.28}_{-0.26}$ & -- & -- \\
$Z$ [solar] & $0.33^{+0.03}_{-0.02}$ & --  & -- \\
$\Delta cz$ [km s$^{-1}$]  & $-510^{+32}_{-32}$  & $-39^{+95}_{-80}$ & $-1274^{+101}_{-106}$ \\
$\sigma_{cz}$ [km s$^{-1}$]  & $95^{+44}_{-65}$ & -- & --\\
{\it Norm} [$\times 10^{-3}$~cm$^{-5}$]  & $1.46^{+0.14}_{-0.14}$ & $0.56^{+0.13}_{-0.13}$ & $0.18^{+0.07}_{-0.07}$ \\ 
{C-stat}/{d.o.f} & \multicolumn{3}{c}{$16748/15990$}   \\
\hline\noalign{\vskip3pt} 
\end{tabular} 
}
\begin{tabnote}
\hangindent6pt\noindent
\hbox to6pt{\footnotemark[$*$]\hss}\unskip%
All errors are at 1$\sigma$ level.
$1$: Metal abundance in solar unit, based on \texttt{LPGS} 
\color{black} table \color{black} on Xspec \citep{2009LanB...4B..712L}.~$2$: 
 \color{black} Line-of-sight velocity, \color{black} relative to those of the optical mean, 6995 km s$^{-1}$. ~$3$: 
\color{black} Velocity dispersion.
\color{black} 
~$4$: The {\tt bapec} normalization ({\it Norm}) is given in $\frac{10^{-14}}{4\pi \left [D_A(1+z) \right ]^{2}}\int n_{e}n_{H}dV$.
\end{tabnote}
\label{table:1ktfit}
\end{table}


\subsection{Two components fit}

Given the long exposure time of \texttt{Center} (398 ks), we divided the Resolve $6\times 6$ array into four quadrants, namely, NW, northeast, southeast and southwest. As noted in Paper I, the NW-quadrant exhibits a positive residual on the low-energy side of the \color{black} Fe He-$\alpha$ \color{black} \textit{w}-line (Figure~\ref{fig:spec_quadNW}, top panel). 
We therefore added a second \texttt{bapec} component with free redshift and normalization, while linking other parameters to those of the first component (hereafter, \color{black} two components  (or 2 comp)  \color{black} model). 
The fitted parameters are shown \color{black} in the middle third of Table~\ref{table:1ktfit}. \color{black}
The normalization of the second component remains non-zero at the 90\% confidence level.
Interestingly, its bulk velocity 
 ($\Delta cz = 5^{+101}_{-94}$~km~s$^{-1}$) 
is consistent with the cluster optical mean and similar to that of NGC 4874 ($\Delta cz = 
172$~km~s$^{-1}$).

\color{black}
The confidence levels reported 
above are estimated using {\bf error} command of XSPEC, based on delta-chi-squared statistics. 
We then used the Akaike information criterion (AIC; \cite{1100705}) and the Bayesian information criterion (BIC; \cite{1978AnSta...6..461S}) to compare the fitting improvements from one component to two components modeling.
The C-stat value improves from 16764.4 (number of bin; $n = 15999$, number of parameter; $k = 5$) to 16755.2 ($k = 7$), which gives $\Delta {\rm AIC} = -5.1$ clearly favoring the latter model. On the other hand, $\Delta {\rm BIC}$ is calculated to be $+10.2$, which is strongly favoring the former simpler model.
Considering the fact that the \texttt{Center} and \texttt{South} average ICM velocity differs by $\sim 300$~km~s$^{-1}$, the secondary component is likely to exist also within a singe observation. Therefore, we think the $\Delta {\rm AIC}$ improvement is important here. 


To assess the significance of the normalization of the second component in a different manner, we generated 300 simulated spectra based on the best-fit one component  model and refitted them with the two component model, fixing the redshift of the second component to the best-fit value $\Delta cz = 5$~km~s$^{-1}$. 
For the second component, only 4 cases of the 300 simulated fits  
\color{black}
(1.3\%) 
\color{black}
exceeded the observed normalization. 
In other words, it is significant in better than 
\color{black}
98\% 
\color{black}
confidence level.

Applying the \color{black} two components \color{black}  model to the full-array \texttt{Center} spectrum also suggests a second redshifted component 
($\Delta cz = -6^{+229}_{-151}$~km~s$^{-1}$). Results are summarized in 
\color{black}
the middile third of 
\color{black}
Table~\ref{table:1ktfit}
and Figure~\ref{fig:spec1} (a) and (c). 
It is significant at the 90\% confidence level.
\color{black}
The C-stat value improves from 17345 ($n = 15999$, $k = 5$) to 17338 ($k = 7$), which gives $\Delta {\rm AIC} = -3$ slightly favoring the latter model (and $\Delta {\rm BIC} = +12.6$ strongly favoring the former).
As an independent check, we generating 300 fake spectra based on the one component model, and fitted with the two components model with fixed second component redshift. Only one case (0.3\%) gave a normalization exceeding that of the observed value. In other words, it is significant in better than $99$\% confidence level.

\color{black}

For the remaining three quadrants, the residuals are less pronounced, and therefore we fixed the second \texttt{bapec} component redshift to that of NGC 4874, and verified the significance of the secondary component. 
However, for the three remaining quadrants, the secondary \texttt{bapec} was insignificant with lower limits consistent with zero at the 90\% confidence level.

\subsection{Third Component in the NW Quadrant}

\begin{figure}
\centering
\includegraphics[width=8cm]{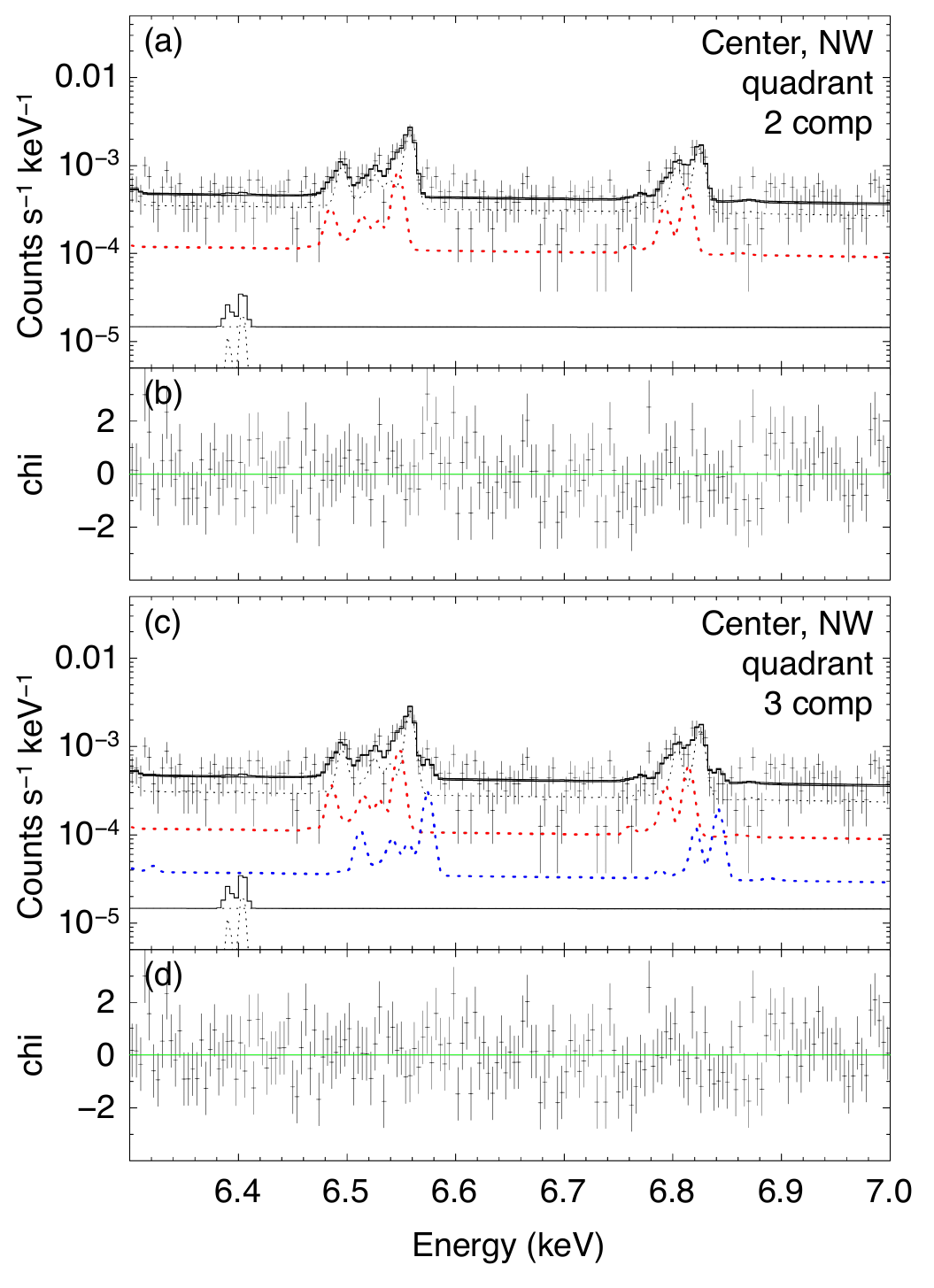}
\caption{
Close-up around  \color{black} Fe-K line  complex \color{black} of the spectra of the NW 
\color{black}
quadrant
\color{black}
of center pointing, fitted with the  two components model (panels (a) and (b)) and the 
\color{black}
components
\color{black}
model (panels (c) and (d)). The secondary redshifted components are shown in red, and the third blueshifted component is shown in blue.
The black component shown in the bottom of panels (a) and (c) represent the NXB model.
{Alt text: Two line graphs, each with an upper part and a lower part. In both parts, the x axis shows the energy from 6.3 to 7 kilo electron volt. In the upper parts, the y axis shows the count rate from 0.00005 to 0.05 counts per second per kilo electron volt. In the lower parts, the y axis shows the residual values from minus 4 to 4.}}
\label{fig:spec_quadNW}
\end{figure}

Inspection of the Center NW-quadrant spectra shown at Figure~\ref{fig:spec_quadNW} (a) and (b) reveals an additional residual on the high-energy side of the \textit{w}-line, 
\color{black}
while the other three quadrant spectra do not show such residuals.
The residual on the former spectra could be 
\color{black}
the ``blueshifted tail'' predicted by \citet{2003AstL...29..791I}. We therefore introduced a third component with an additional \texttt{bapec} model (hereafter
\color{black}
the three components (or 3 comp)
\color{black}
model). The fitted parameters are shown in 
\color{black}
the bottom third of Table~\ref{table:1ktfit} 
\color{black}
and the spectrum and residuals are shown in Figure~\ref{fig:spec_quadNW} (c) and (d). 
Both the second and third components have non-zero normalizations at confidence levels of 90\%. 

\color{black}
We generated 300 simulated spectra based on the best-fit 
one component 
model and refitted them with the three components model, fixing the redshifts of the second and third components to their best-fit values. 
\color{black}
For the second component, only 15 (5\%) cases of the 300 simulated fits exceeded the observed normalization, and for the third component only 10 (3.3\%). In other words, the second component is significant in $95$\% confidence level, and the third component in $\sim 96$\% confidence level. 

\color{black}
We also noted the lack of ``blueshifted tail'' signal in the Fe Ly$\alpha$1 line. By performing yet another 300 fake spectra with the best-fit three components model, we found that 8\% of the simulations show photon counts less than those observed within 6.83--6.85~keV, where the deficit exists in the data. In other words, the deficit in the blueshift region of Fe Ly$\alpha$1 line is not statistically significant.
There is also a possibility that the blueshifted component has a lower temperature with smaller intensity in Fe Ly$\alpha$1 line.
Although it is beyond the scope of this paper, temperature sorted velocity distribution in the Coma ICM is an interesting topic for future work.
\color{black}

\section{Multiple redshift component model fit}
\label{S_51comp}

\subsection{Methodology}

Detection of the second and third ICM components in the \texttt{Center} NW-quadrant spectra indicates that the velocity distribution 
is not entirely described by a simple Gaussian \color{black} broadening. \color{black} In other words, non-Gaussianity is present, as expected, and fitting models assuming Gaussian-broadened velocity dispersion may not be always valid in a high-statistics data like those of the Coma cluster.

To address this issue with minimal 
assumptions, we adopted a less-biased fitting procedure to search for differential contributions of ICM components along the redshift axis. Specifically, we applied 
51 ICM components, represented by the bapec model, spaced at 3~eV intervals (hereafter the 3~eV stepped fitting). This step size is slightly smaller than the Resolve energy resolution ($\sim 4.5$~eV FWHM).
Actually, because of the thermal broadening of about 
\color{black}
$2.5 \times \sqrt{\frac{kT}{8~{\rm keV}}}$~eV 
\color{black}
(already included in \texttt{bapec} model), total line width is 
\color{black}
$\sim$ 7.5~eV 
\color{black}
without any velocity dispersion. The 3~eV step we adopted is smaller than this value to not overlook the 
observed line broadening.
\color{black} 
Because this 3 eV stepped fitting is computationally heavy, all the parameters must be optimized not keep it feasible. 
If we make this number even smaller, the covariance of normalizations in the fitting process will significantly increase, and the computational cost rises rapidly. We therefore keep the step at 3~eV in this work. 
\color{black}
Each component was modeled with \texttt{bapec}, fixing the velocity dispersion to zero while linking temperature and abundance across all components. Normalizations were freed to vary, but components contributing to less than  1\% of the main peak 
of the 1kT fit
were set to zero during the fittings, to converge the process.
In this approach, the 3~eV shifted components have a fixed ``origin'' in redshift, and this origin will cause systematic effect. To assess it, we repeated the fitting with the origin shifted by $\pm 1$~eV, producing three sets of results.
\color{black} 
To justify this number, we increased the shift increments from 1~eV to 0.3~eV, but the detected peak normalization changed only less than 1\%, and therefore we conclude 1~eV shift is sufficient for our analysis.

\color{black}

\begin{figure}[h]
\centering
\includegraphics[width=8cm]{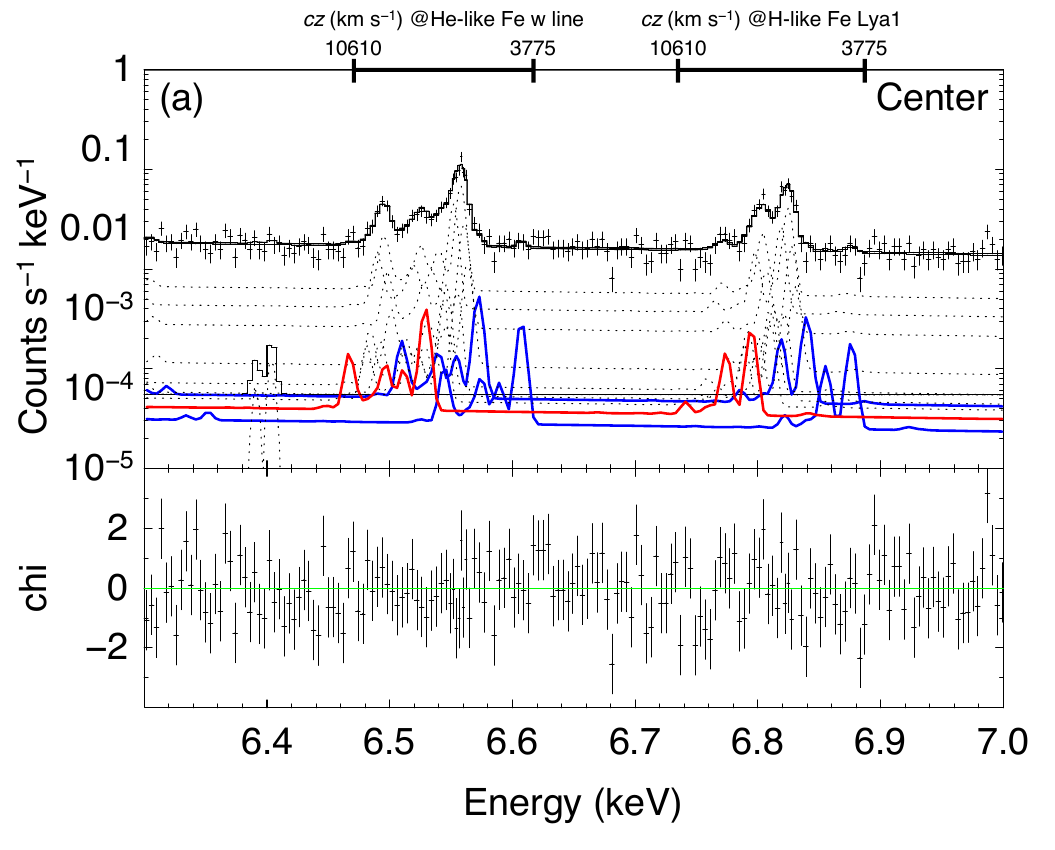} 
\hfill
\includegraphics[width=8cm]{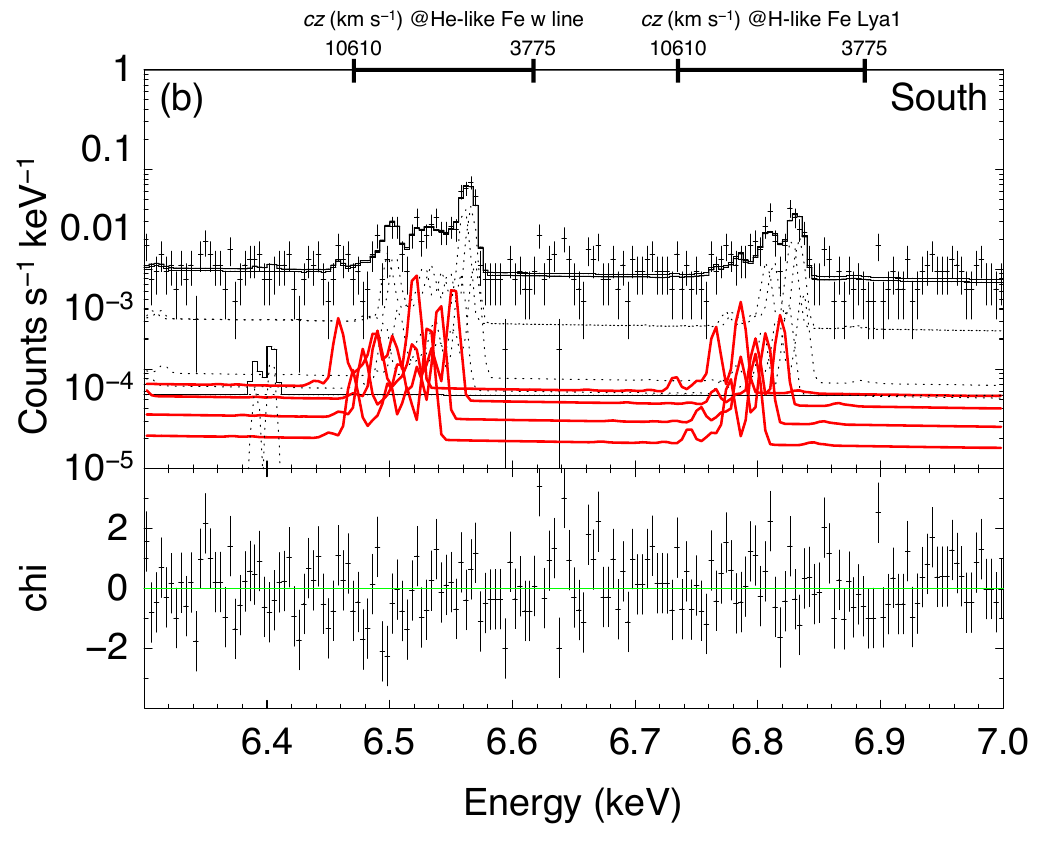} 
\caption{Results of 3~eV stepped fitting to (a) \texttt{Center} and (b) \texttt{South} spectra. The red and blue solid lines indicate the red/blueshifted components in each observation, respectively. In the \texttt{South} spectra, no blueshifted component was needed. See text for detail. The x axis at the top of each panel shows the range of 
\color{black} line-of-sight velocities \color{black} 
at the \color{black} Fe He-$\alpha$ \color{black} \textit{w}-line and the 
Fe Ly$\alpha$1 line. {Alt text: Two line graphs. Panels (a) and (b) each consist of an upper part and a lower part. In both panels, the bottom x axis shows the energy from 6.3 to 7 kilo electron volt. The range of the top x axis is 3775 to 10610 kilo meter per second. In the upper parts, the y axis shows the count rate from 0.00001 to 1 counts per second per kilo electron volt. In the lower parts, the y axis shows the residuals from minus 4 to 4.}}
\label{fig:3eVfitting_0}
\end{figure}


\subsection{Main peak in the redshift distribution}

\begin{figure*}[hbtp]
\centering
\begin{minipage}[b]{0.49\textwidth}
\centering
\includegraphics[width=0.95\textwidth]{
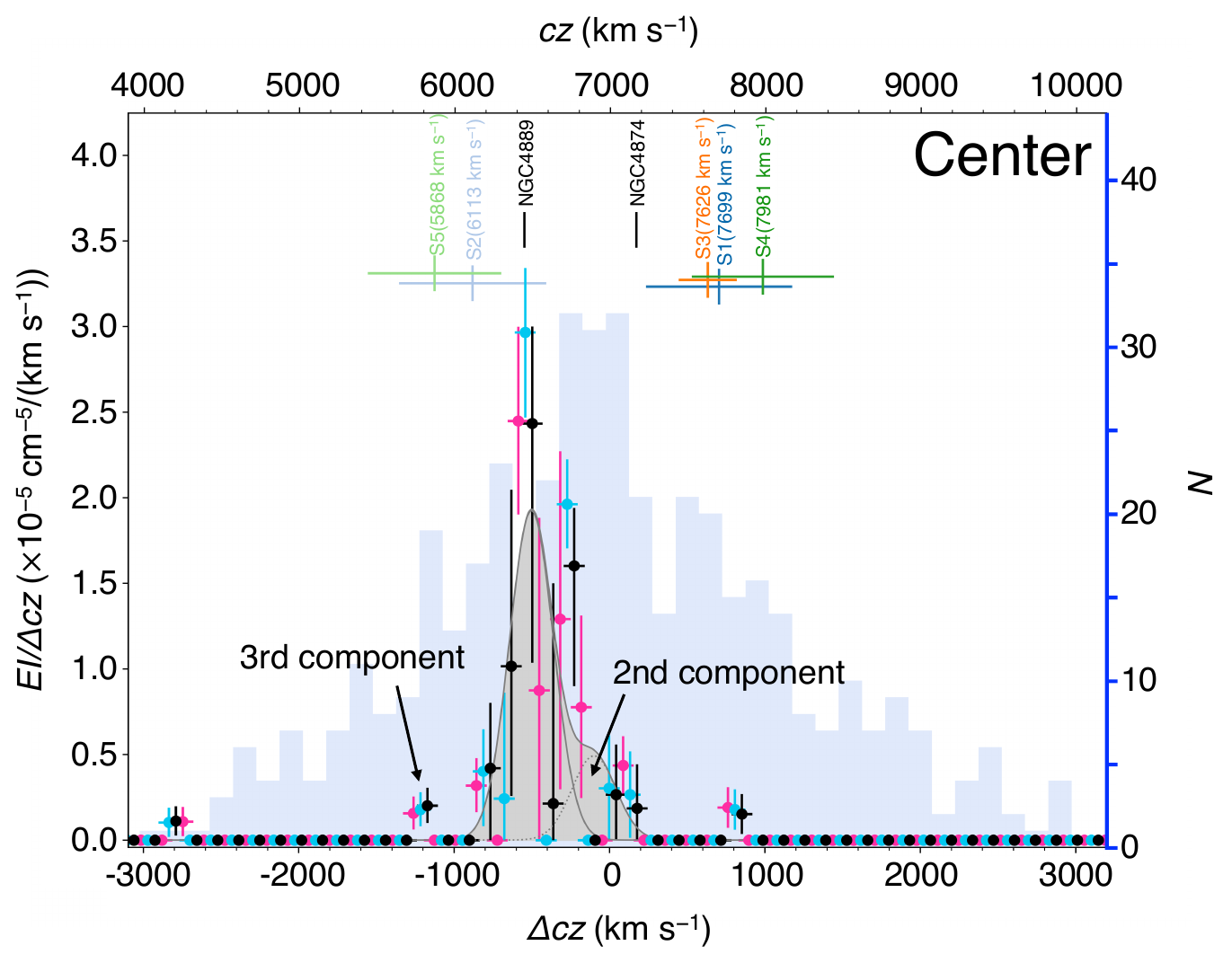}
\end{minipage}
\centering
\begin{minipage}[b]{0.49\textwidth}
\centering
\includegraphics[width=0.95\textwidth]{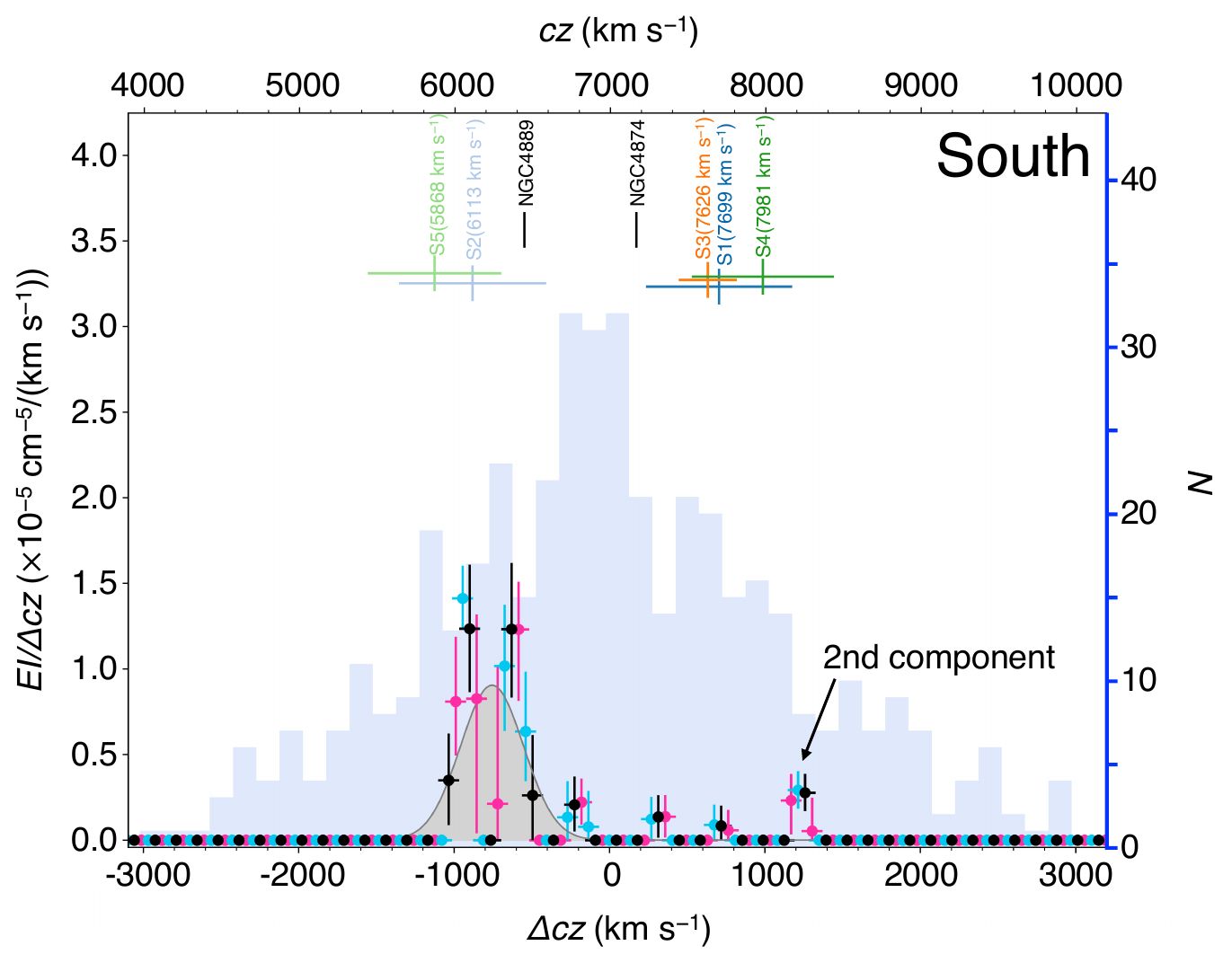}
\end{minipage}
\caption{Results of 3~eV stepped fitting to the Center (left) and South (right) spectra, shown as a cross with dots. Horizontal axis is \color{black} $\Delta cz$. \color{black}
Results with $+$ and $-1$~eV shift of the redshift is also shown as pink and sky-blue dots, respectively. Overlaid in gray is the results of \color{black} two components  \color{black}  (\texttt{Center}) and \color{black} one component  \color{black}  (\texttt{South}) fitting with velocity dispersion in Gaussian functions. In the \texttt{Center} plot, there are the main component, the second component (redshifted) and the third component (blueshifted).
In the \texttt{South} plot, there are the main component and the second redshifted component. See table~\ref{table:redshifts_4} for detail.
The light-blue histogram overlaid are the velocities of the Coma cluster member galaxies within $r<20'$ from the X-ray center (\cite{2025ApJ...985L..20X}, Figure 7), and the colored crosses on top represents the $cz$ of the five optical subgroups with their velocity dispersions. 
{Alt text: Two distribution plots. In both the left and right panels, the range of the horizontal axis is approximately 
\color{black}
$-3000$ to $3000$ kilo meter per second,
relative to the cluster mean.
\color{black}
Vertical axis represents the emission integral normalized per unit 
line-of-sight velocity 
bin, with units of per cubic centimeter to the minus 5 power per three kilo meter per second. 
A light-blue histogram overlaid with two distribution plots. The top x axis shows 
\color{black}
line-of-sight velocity of 3900 to 10200 kilo meter per second.
\color{black}
The right y axis corresponds to the histogram and indicates the number of galaxies from 0 to 44.} }
\label{fig:3eVfitting_opt}
\end{figure*}

Because this method requires high 
statistics, here we focus on
full-array spectra of \texttt{Center} and \texttt{South}.
Figure~\ref{fig:3eVfitting_0} \color{black} shows  \color{black} the results of the 3 eV stepped fitting.
In Figure~\ref{fig:3eVfitting_opt}, we plotted the redshift distribution from 
our fit. For comparison, fitted results from Gaussian-based fitting of \texttt{Center} \color{black} two components  \color{black}  model, and \texttt{South} \color{black} one component  \color{black}  model are also presented.
The main peaks are reproduced by several non-broadened components, as expected: a velocity dispersion of
200--300~km~s$^{-1}$ (1 $\sigma$) corresponds to 10--15~eV in FWHM at 6.5~keV, requiring three to five 3~eV components to represent the peak.
Error-bars of the normalizations of 3-eV-stepped components are not independent among neighboring ones, and therefore inevitably overestimated. 

In the \texttt{Center} spectrum, the 3~eV stepped-fitting result include two points (in default base, shown in black) near $cz \sim 7000$~km~s$^{-1}$, corresponding to the second component identified in the last subsection.  
In the ``default +1eV shifted'' plot (pink data in Figure~5), the redshifted component is expressed by a single component (i.e., the neighboring two components have zero normalization), with a normalization significance of $2.6~\sigma$. When such ``single component fit'' was obtained in the $0,~ \pm 1$~eV origin shift results, we regard it as the ``plausible significance'' of the component.

\subsection{Sub-peak candidates}
\label{sec:subpeak}

In the \texttt{Center} full-array data, another point around $cz \sim 5850$~km~s$^{-1}$ reflects the third component, which was detected in the NW quadrant (but not significantly strong in the full-array data). Also, yet another blueshifted component candidate (around $cz \sim 4200$~km~s$^{-1}$) and a redshifted component candidate (around $cz \sim 7800$~km~s$^{-1}$) were present, with ``plausible 
significances'' of 1.3$\sigma$ and 1.6$\sigma$.

In the \texttt{South} data, the distribution shows a broadened main peak, and a few redshifted components
at $cz \sim 7300,~\sim 7700$, and $\sim 8250$~km~s$^{-1}$, with ``plausible significances'' of 1.1$\sigma$, 1.0$\sigma$ and 2.7$\sigma$, respectively.

These significance values, however, must be interpreted with caution. 
For example, if there are 20 independent trials, 
a peak below $2.55\sigma$ (equivalent to a probability of $1-0.9946$) corresponds to a $>10$\% chance of arising from statistical fluctuation (because $1-0.9946^{20} = 0.10$). 
This is the so-called ``look-elsewhere effect''.
However, $\sigma$ vs probability relation does not always follow the Gaussian distribution, and therefore additional analyses based on simulations are needed.

For simplicity, we decide to ignore sub-peak candidates with significances smaller than $2.0\sigma$. In this case, there are only two candidates: ``the \texttt{Center} blue-wing'' at $cz \sim 5850$~km~s$^{-1}$ (with 2.0$\sigma$), and ``the \texttt{South} red-wing'' feature at $cz \sim 
8250$~km~s$^{-1}$ (with 2.7$\sigma$).
To estimate their significance, we made another \color{black} 100 
\color{black} fake spectra (using XSPEC \texttt{fakeit} command) with Gaussian broadening using the best-fit \color{black} two components  \color{black} model of the \texttt{Center} data, and the \color{black} one component  \color{black}  model of the \texttt{South} (see Table~\ref{table:1ktfit}).
Then we applied the same ``3~eV stepped fitting'' procedure to these simulated spectra. By this approach, we can estimate how pure statistical fluctuation can lead to false 
detection of redshifted ICM components. 

In the \texttt{Center} case, within the ``blueshifted''  range of $3775\leq cz\leq 5980~\rm{km~s^{-1}}$ (16~shifted values), there are 8 
cases exceeding the normalization of the fitted value, among the $100$ (fake simulation number) times $3$ (the $0,~\pm 1$~eV origin shifts) cases. This gives $8/(100\times 3) = 2.7$\% probability. In other words, the real data fitting to ``the \texttt{Center} blue-wing'' has a null-hypothesis probability (NHP) of 
3\% for the detection.

Similarly, in the \texttt{South} ``redshifted'' range of $6745\leq cz\leq 10610~\rm{km~s^{-1}}$ (29~shifted values), there are 16 cases exceeding the normalization of the fitted value 
of ``the \texttt{South} red-wing''. 
This gives $16/(100\times 3) = 5.3$\% probability. In other words, the real data fitting to the ``the \texttt{South} red-wing'' has a NHP of 
5\% for the detection.
In the ``the \texttt{South} red-wing'' case, however, we need to be careful about the ICM emission model systematics. 
XRISM reports indicate that certain \color{black} Fe He-$\alpha$ 
line components, $x,~y,~z$ may be underestimated, producing positive residuals (e.g. \cite{10.1093/pasj/psaf089}). 
For example, if excess in $y$ line is misunderstood as a redshifed $w$ line, the velocity difference will be fitted as
\color{black} $\sim 1481$~km~s$^{-1}$.
\color{black}
The strongest line (\textit{w}) of ``the \texttt{South} red-wing'' candidate does not coincide with these transitions and accounts for $\sim 30$\% of the flux at its energy. We therefore conclude that this residual is unlikely to be a modeling artifact and regard the component as marginally detected at $95$\% confidence level.

\color{black}
Although handling the blue and redshift regions in a more flat manner is desirable, they differ significantly. In addition to the systematic errors mentioned above, the redshift region has larger statistical errors as well, because these satellite lines are stronger than the Bremsstrahlung continuum which is the only matter in the blueshift region. Therefore, we handled them separately here.
\color{black}

\section{Discussion}

\subsection{Redshift comparison between ICM and galaxies}

Figure~\ref{fig:3eVfitting_opt} compares the ICM redshift distribution with that of Coma cluster member galaxies. As summarized in \color{black} Table~\ref{table:redshifts_4}, \color{black} the primary ICM component in \texttt{Center} is blueshifted by
$\Delta cz= -498^{+31}_{-33}$~km~s$^{-1}$ 
relative to the cluster mean and has the velocity similar to one of the two main cluster galaxies, NGC 4889 ($\Delta cz= -549$~km~s$^{-1}$). The second component ($\Delta cz= -6^{+229}_{-151}$~km~s$^{-1}$) 
is consistent with the cluster galaxy mean and close to the velocity of the other main galaxy, NGC 4874 ($\Delta cz= 
172$~km~s$^{-1}$). The third component in the NW-quadrant ($\Delta cz= -1274^{+101}_{-106}$~km~s$^{-1}$) is also present, \color{black} while the other two major components are consistent with the two components in the \texttt{Center} whole spectra.  
In the \texttt{South} \color{black} pointing, the main ICM component is blueshifted by $\Delta cz \sim -750$~km~s$^{-1}$, and \color{black} the second component candidate identified in the 3~eV stepped-fitting analysis at $\Delta cz \sim 1250$~km~s$^{-1}$ (or $cz \sim 8250$~km~s$^{-1}$) exists.  
\color{black} 

\begin{table}[htbp]
\begin{center}
\tbl{\color{black} Summary of redshifts of selected ICM components and optically selected \color{black} galaxy groups \label{table:redshifts_4} }{%
\begin{tabular}{lcll}  
\hline\noalign{\vskip3pt} 
Region & Component & $\Delta cz$ $^1$ & Norm$^2$  \\
\hline
Center & 1st & \color{black} $-481^{+25}_{-38}$ \color{black} & $7.8^{+0.6}_{-1.0}\times 10^{-3}$ \\
 & 2nd & $-6^{+229}_{-151}$ & $1.1^{+1.0}_{-0.6}\times 10^{-3}$ \\
Center-NW & 3rd$^3$ & $-1274^{+101}_{-106}$ & $0.2^{+0.1}_{-0.1}\times 10^{-3}$ \\
\hline
South & 1st & $-744^{+22}_{-23}$ & $5.1^{+0.1}_{-0.1}\times 10^{-3}$ \\
 & 2nd$^4$ &  $\sim 1250$ &  $0.4^{+0.2}_{-0.1}\times 10^{-3}$ \\
\hline\noalign{\vskip3pt} 
\end{tabular}}
\begin{tabular}{lccccc}  
\hline\noalign{\vskip3pt} 
\hline\noalign{\vskip3pt} 
Group ID$^5$ & S1 & S2 & S3 & S4 & S5 \\
\hline\noalign{\vskip3pt} 
$\Delta cz$ $^1$& 
~700~ & $-886$ & ~627~ & ~982~ & $-1131$ \\
$\sigma cz$ & 
~924~ & ~932~ & ~358~ & ~899~ & ~843~ \\
\hline\noalign{\vskip3pt} 
\end{tabular}
\begin{tabnote}
\hangindent6pt\noindent
\hbox to6pt{\footnotemark[$*$]\hss}\unskip%
1: 
\color{black} Line-of-sight velocity \color{black}
in km s$^{-1}$, relative to those of the optical mean \color{black} (6995 \color{black} km s$^{-1}$). ~2: \color{black}
The {\tt bapec} normalization. ~3: Derived from the three components model fitting \color{black} to the NW quadrant spectra. ~4: The sub-peak identified in \S~\ref{sec:subpeak} ~5: Galaxy group ID from \cite{2021AandA...650A..76H}. S1 includes NGC~4874 (the west BCG) and S2 includes NGC~4889 (the east BCG). 
\end{tabnote}
\end{center}
\end{table}

To compare these values with those of optical substructures, we examined galaxy groups identified by \citet{2021AandA...650A..76H}.
\color{black} As already noted in Figure~\ref{fig:image},
\color{black}
NGC~4889 is located $\sim 7'.5$ east of NGC~4874, and the \texttt{Center} observation (located at the X-ray peak) covers $\sim 3'\times3'$ region east of NGC~4874. \texttt{South} observation is $\sim 6'$ south of \texttt{Center}.
We therefore focused on the 5 groups located within $10'$ from \texttt{Center}, and also listed their redshifts in Table~\ref{table:redshifts_4}.
Two groups (S2 including NGC~4889 and 4894, and S5 including NGC~4869) are strongly blueshifted ($\Delta cz = -886$ and $-1131$~km~s$^{-1}$), while three groups (S1, S3, S4) are redshifted, including S1 ($\Delta cz = 700$~km~s$^{-1}$) including NGC 4874. The redshift of S4 ($\Delta cz = 982$~km~s$^{-1}$) is similar to that of the \texttt{South} red-wing candidate.
There is no galaxy group with redshift comparable to the cluster mean among the five selected groups. 
For comparison, redshift values and velocity dispersions of the five groups are plotted as crosses on top of Figure~\ref{fig:3eVfitting_opt}.

Although the \texttt{Center} NW-quadrant third component and the \texttt{South} red-wing feature are marginal detections (with 3\% and 5\% NHP, respectively), their coincidence with optically identified subgroups are interesting. One possibility is galaxy\UTF{2013}ICM interactions (\cite{2020AandA...638A.138G}), which can involve kinetic energy transfer from galaxies to the ICM, generating turbulence and forming wakes detached from the global velocity field. Another 
possible explanation 
is the merger geometry. The S1 group (including NGC~4874) is located in the middle of the biggest gravitational sub-halo (\cite{Okabe_2014,Kang_2025}) and the ICM will be accumulated around it. 
The \texttt{Center} second and the third components can be caused by the rotation induced by interaction with the S2 group (including NGC~4889). On the other hand, the \texttt{South} red-wing can be the still infalling ICM associated with the S1, S3 and S4 groups. 
Given the limited information of these features within the \color{black}
existing
\color{black}
data, however, further XRISM observations will be needed, and detailed interpretation is beyond the scope of this paper.

\subsection{Non-thermal energies in the ICM}


Our analysis reveals at least two strong 
(and the third weaker) ICM components in \texttt{Center} and one (and another weaker) one(s) in \texttt{South}. Based on the bulk velocity and velocity dispersions associated with these components, here we estimate the non-thermal energy densities. For simplicity, we ignore the \texttt{Center} third component and the \texttt{South} red-wing candidate, and utilize the numbers \color{black} of the two component model for \texttt{Center} and the one component model for \texttt{South}, as summarized in Table~\ref{table:redshifts_4}.
 \color{black}

The cluster central electron density is estimated as $n_e=3.42\pm 0.05 \times 10^{-3}$~cm$^{-3}$ (\cite{1992AandA...259L..31B}, scaling $H_{50}$ to $H_{70}$). 
\color{black} 
Because our Resolve data only covers a small portion of the cluster center, we relied on the existing results from wide-field ROSAT/PSPC in this analysis.
 \color{black}
Total particle density and mass density are  $n_{\rm total}= n_e \times 1.932 = 6.61 \pm 0.09 \times 10^{-3}$~cm$^{-3}$ and $\rho= m_p (n_e\times (0.864+0.068\times 4))=6.50 \pm 0.09 \times 10^{-27}$~g~cm$^{-3}$. 
For \texttt{South}, we scale these values by 0.76 based on the square root of the normalization as a simple approximation. 

Calculating the bulk energy of the two components in \texttt{Center} is not easy. For simplicity, we estimated their masses to be proportional to their $Norm$. In other words, we adopted the simplest assumption that the two components have the same density, temperature, velocity dispersion (and hence the same pressure), but filling separate volumes with their volume filling factors proportional to their normalizations.
For the bulk motion, we consider two reference frames: the cluster optical mean ($cz =  6995$~km~s$^{-1}$ or $\Delta cz= 0$~km~s$^{-1}$ ) and the primary ICM component in \texttt{Center} ($cz = 6497$~km~s$^{-1}$ or $\Delta cz= -498$~km~s$^{-1}$ ).

Thermal energy density is $u_{\rm th} =\frac{3}{2} n_{\rm total} kT$ erg cm$^{-3}$, turbulent energy density is 
\color{black}
$u_{\rm turb.3D} =\frac{1}{2} \rho \sigma_{3D}^2=3/2 \rho \sigma_v^2$ erg cm$^{-3}$, 
\color{black}
and one-dimensional bulk kinetic energy density is $u_{\rm bulk.1D}=\frac{1}{2} \rho v_{\rm bulk}^2$~erg~cm$^{-3}$. 
\color{black}
Values derived from the line-of-sight velocity relative to the optical mean are denoted as $u^{\rm ave}_{\rm bulk.1D}$, whereas those relative to the first ICM component in the Center region are denoted as $u^{\rm ICM}_{\rm bulk.1D}$.
\color{black}
Here, motions on the sky-plane is ignored because we have no information on them. 
\color{black}
We also investigated how the best fit bulk velocity and velocity dispersion can differ if the thermal model temperature is in-correct, for example, by selection of arf and/or mixture of ICM components. In short, even if the best-fit temperature differs by $\pm 1$~keV, these velocity values differ less than a few km~s$^{-1}$, significantly smaller than their statistical errors.
\color{black}

\begin{table}[htbp]
\tbl{ICM thermal and non-thermal energy densities}
{\begin{tabular}{l|l|l|l}  
\hline
Region & \multicolumn{2}{c|}{Center} & South \\
component & 1st  & 2nd & 1st\\
\hline
\hline
$u_{\rm B}$ [eV cm$^{-3}$] & \multicolumn{2}{c|}{$\sim 0.55$} & $\sim 0.42$ \\
$u_{{\rm CR}e}$ [eV cm$^{-3}$] $^1$ & \multicolumn{2}{c|}{$\sim 7.0\times 10^{-4}$} & $\sim 5.9\times 10^{-4}$  \\
$u_{\rm th}$ [eV cm$^{-3}$]& \multicolumn{2}{c|}{$82.9\pm 1.8$} & $55.9^{+1.9}_{-2.0}$ \\
$u_{\rm turb.3D}$ [eV cm$^{-3}$] & \multicolumn{2}{c|}{$1.9^{+0.6}_{-0.7}$} & $2.2^{+0.5}_{-0.4}$ \\
\hline\noalign{\vskip3pt} 
$u_{\rm bulk.1D}^{\rm ave}$ [eV cm$^{-3}$]$^2$  & $4.1^{+0.4}_{-0.6}$ & \color{black} $<0.10$ & $8.5^{+0.5}_{-0.5}$ \\
$u_{\rm bulk.1D}^{\rm ICM}$ [eV cm$^{-3}$]$^3$  & \color{black} -- & $0.6^{+0.6}_{-0.4}$ & $1.1^{+0.3}_{-0.4}$ \\
\hline\noalign{\vskip3pt} 
\end{tabular}}
\begin{tabnote}
\hangindent6pt\noindent
\hbox to6pt{\footnotemark[$*$]\hss}\unskip%
All values are in eV cm$^{-3}$. ~1: Relativistic electron energy density was integrated over $10^2<\gamma<10^5$. ~2: Bulk motion (1D) energy relative to optical mean.~3: Bulk motion (1D) energy relative to the 1st \color{black} ICM \color{black} component of the Center.
\label{table:energy}
\end{tabnote}
\end{table}

Magnetic field estimates from rotation measure studies are $B_0 \sim 4.7~\mu$G at \texttt{Center} and $\sim 4.1~\mu$G at \texttt{South} ($6'$ offset) (see Figure~20 of \cite{2010AandA...513A..30B}).
 Their energy densities are $u_B=\frac{B^2}{8 \pi} \sim 0.55$~eV~cm$^{-3}$ and $\sim 0.42$~eV~cm$^{-3}$, respectively. 
With this magnetic field, 
the radio halo flux at 1.4 GHz ($\sim 80$~mJy per beam at 1.4 GHz; \cite{1997AandA...321...55D}) 
implies a relativistic electron energy density 
\color{black}
$u_{{\rm CR}e} \sim 7\times 10^{-4}$~eV cm$^{-3}$, 
\color{black}
negligible compared to thermal and kinetic components. 

\color{black}
For a magnetic field strength of $4.7~\mu$G and a plasma density of $\sim 6.50 \times 10^{-27}$~g~cm$^{-3}$ at \texttt{Center},
the Alfv\'en velocity is $v_{\rm A} \simeq 165$~km~s$^{-1}$. The observed one-dimensional velocity dispersion of $\sigma_{cz} = 177$~km~s$^{-1}$ from the two components model
gives an isotropic turbulence $\sigma_{cz.{\rm 3D}} = 307$~km~s$^{-1}$, corresponding to an Alfv\'en Mach number of $M_{\rm A.3D}=1.9$. Thus, the ICM motions are approximately trans- to mildly super-Alfv\'ennic. If a significant fraction of the velocity dispersion originates from overlapping bulk motions, local (or small-scale) Alfv\'en Mach number will be even less.
\color{black}

Overall energy densities are summarized in Table~\ref{table:energy}, including the thermal ones.
In general, ICM bulk motion kinetic energy density is similar to those of turbulence.
Interestingly, the magnetic field energy density is a fraction of the turbulent energy. Similar to those observed in the Milky way galaxy, this relation supports the idea that turbulence is generating the magnetic field. Note that the measured velocity dispersion represents an upper limit on turbulence, as it includes line-of-sight bulk motions. Therefore, difference between the two values can be smaller in reality.

The turbulence dissipation rate is $Q_{\rm turb} \sim 5 \rho v_{\rm turb.1D}^3/l_t$. Here  $l_t$ is the length scale at which velocity is measured (\cite{2014Natur.515...85Z}). Assuming $l_t = 3'$ (84.6~kpc), and $v_{\rm turb.1D} = 177$~km~s$^{-1}$ (in the \texttt{Center} full-array),
$Q_{\rm turb}$ becomes $\sim 4.3\times 10^{-16}$~eV~cm$^{-3}$~s$^{-1}$. This means typical decay time of the turbulence $\tau = u_{\rm turb.3D}/Q_{\rm turb}$ is $\sim 0.14$~Gyr.
Because the velocity dispersion reflects the line-of-sight integration of 
\color{black}
isotropic turbulent and unresolved bulk motions of the 
\color{black}
ICM, effective $l_t$ must be larger, say the core radius, $\sim 500$~kpc, at the largest. In this case, $Q_{\rm turb}$ becomes $\sim 1/6$, and the $\tau$ will be about a Gyr.
Synchrotron-emitting electrons at 144 MHz in a  5$~\mu$G magnetic field have 
cooling times of $\sim 0.25$~Gyr (e.g., \cite{1999ApJ...520..529S}), comparable to turbulence decay times. 
Bulk kinetic energy will eventually convert into turbulence, extending its lifetime. A bulk flow of 500~km~s$^{-1}$ requires $\sim 1$~Gyr to traverse 500~kpc, similar to the cluster core size. 
As such, continuous decay of turbulence is needed to support the $\sim 500$~kpc wide radio halo.
Including the NW third component and South red-wing candidate will add a little more energy, although the overal picture does not alter a lot.

\section{Conclusions}

By analyzing XRISM/Resolve 
\color{black}
data
\color{black}
of the Coma cluster’s \texttt{Center} and \texttt{South} regions, we confirm the presence of non-Gaussian features in the \color{black} Fe-K line  complex \color{black} profiles that indicate variations of the ICM bulk velocity on the line of sight.
In the Center pointing, we detect a second redshifted component 
with $\Delta cz = -6^{+229}_{-151}$~km~s$^{-1}$.
In addition, in the NW quadrant we find a third component with $\Delta cz = -1274^{+101}_{-106}$~km~s$^{-1}$s 
 (Table~\ref{table:1ktfit}).

Using a non-biased multi-component fitting approach with 3 eV steps to the two full-array spectra, we identify two additional candidates in \texttt{Center} ($cz \sim 5850$~km~s$^{-1}$, or $\Delta cz \sim -1150$~km~s$^{-1}$), and \texttt{South} ($cz \sim 8250$~km~s$^{-1}$, or $\Delta cz \sim 1250$~km~s$^{-1}$). The former is consistent with the third component found in the \texttt{Center}-NW-quadrant 
\color{black}
three components model fitting.
\color{black}
\color{black}
Monte-Carlo simulations indicate the presence of these features at the 97\% confidence level for Center and $\sim 95$\% for South.
\color{black}
The primary, second, and third components in \texttt{Center}, as well as the main and red-wing candidate components in \texttt{South}, show possible associations with optically identified galaxy groups within $10'$ of the cluster center.

We estimate that the magnetic field energy density is only a fraction of the ICM turbulence and bulk-motion kinetic energy densities, which in turn are comparable to each other. The estimated turbulence decay timescale  $\tau = u_{\rm turb}/Q_{\rm turb} \sim 0.14$~Gyr is similar to the synchrotron cooling timescale of relativistic electrons, $\sim 0.25$~Gyr at 144~MHz within $5~\mu$G magnetic field.
If the ICM dissipates its bulk and turbulent kinetic energy into relativistic electrons generating the radio halo, the halo could be sustained for about 1~Gyr.

\begin{ack}
This work was supported by JSPS KAKENHI grant numbers 
JP25K07368 (N.O.), JP20H00157 (K.N.), and JP25K23398 (S.U.), 
N.O. acknowledges partial support by the Organization for the Promotion of Gender Equality at Nara Women's University.
S.U. acknowledges support by Program for Forming Japan's Peak Research Universities (J-PEAKS) Grant Number JPJS00420230006. I.Z. acknowledges partial support by NASA grant 80NSSC18K1684 and the Alfred P. Sloan Foundation through the Sloan Research Fellowship.
This work made use of data from the Galaxy Cluster Merger Catalog (http://gcmc.hub.yt).
\end{ack}

\bibliographystyle{pasj} 
\bibliography{pasj} 

\end{document}